\documentstyle[12pt]{article}
\def\zid{1\kern-0.36em\llap~1}

\newcommand{\beq}{\begin{equation}}
\newcommand{\ber}{\begin{eqnarray}}
\newcommand{\eeq}{\end{equation}}
\newcommand{\eer}{\end{eqnarray}}

\begin{document}

\begin{titlepage}
\rightline{[SUNY BING 2/9/04] } \rightline{ hep-ph/0402255}
\vspace{2mm}
\begin{center}
{\bf \hspace{0.1 cm} ON AN INTENSITY-RATIO EQUIVALENCE-THEOREM
\newline
FOR TOP QUARK DECAY }\\
\vspace{2mm} Charles A. Nelson\footnote{Electronic address:
cnelson @ binghamton.edu  } \\ {\it Department of Physics, State
University of New York at Binghamton\\ Binghamton, N.Y.
13902}\\[2mm]
\end{center}


\begin{abstract}

For the $t \rightarrow W^{+} b $ decay mode, an intensity-ratio
equivalence-theorem for two Lorentz-invariant couplings is shown
to be related to symmetries of tWb-transformations. Explicit
tWb-transformations, $A_{+}=M $ $A_{SM}$, $P $ $A_{SM}$, $B $
$A_{SM}$ relate the four standard model's helicity amplitudes,
$A_{SM}\left( \lambda _{W^{+} } ,\lambda _b \right)$, and the
amplitudes $A_{+}\left( \lambda _{W^{+} } ,\lambda _b \right)$ in
the case of an additional $t_R \rightarrow b_L$ weak-moment of
relative strength $\Lambda_{+} =E_W /2 \sim 53 GeV$. Two
commutator plus anti-commutator symmetry algebras are generated
from $ M, P, B$. These transformations enable a characterization
of the associated mass scales.

\end{abstract}

\end{titlepage}

\section{Introduction:}

In this paper, for the $t \rightarrow W^{+} b $ decay mode [1], an
intensity-ratio equivalence-theorem [2] for two Lorentz-invariant
couplings is shown to be related to symmetries of
tWb-transformations, $A_{+}=M $ $A_{SM}$, $P$ $A_{SM}$, $B $
$A_{SM}$, where $ M, P, B$ are explicit $4$x$4$ matrices. These
\newline tWb-transformations relate the standard model's helicity
amplitudes, $A_{SM}\left( \lambda _{W^{+} } ,\lambda _b \right)$,
and the amplitudes $A_{+}\left( \lambda _{W^{+} } ,\lambda _b
\right)$ in the case of an additional $t_R \rightarrow b_L$
weak-moment of relative strength $\Lambda_{+} =E_W /2 \sim 53
GeV$.  Versus the standard model's pure $(V-A)$ coupling, the
additional tensorial coupling can be physically interpreted as
arising due to a large chiral weak-transition moment for the top
quark. $\Lambda_{+}$ is defined by (1) below and the $(+)$
amplitudes' complete coupling is (2). $E_W$ is the energy of the
final W-boson in the decaying top-quark rest frame. The subscripts
$R$ and $L$ respectively denote right and left chirality of the
coupling, that is $( 1 \pm \gamma_5 ) $. $\lambda _{W^{+} }$, $
\lambda _b $ are the helicities of the the emitted W-boson and
b-quark in the top-quark rest frame.   The Jacob-Wick
phase-convention [3] is used in specifying the phases of the
helicity amplitudes and so of these transformations.

Due to rotational invariance, there are four independent $A\left(
\lambda _{W^{+} } ,\lambda _b \right)$ amplitudes for the most
general Lorentz coupling. Stage-two spin-correlation functions
were derived and studied as a basis for complete measurements of
the helicity parameters for $t \rightarrow W^{+} b $ decay as
tests with respect to the most general Lorentz coupling [4,5].
Such tests are possible at the Tevatron [1], at the LHC [6], and
at a NLC [7]. In this paper, a subset of the most general Lorentz
coupling is considered in which the subscript ``$i$" identifies
the amplitude's associated coupling: ``$i=$ SM" for the pure
$(V-A)$ coupling, ``$i=(f_M + f_E)$" for only the additional $t_R
\rightarrow b_L$ tensorial coupling, and ``$i=(+)$" for $(V-A) +
(f_M + f_E)$ with a top-quark chiral weak-transition moment of
relative strength $\Lambda_{+} =E_W /2$ versus $g_L$. The Lorentz
coupling involving both the SM's $(V-A)$ coupling and an
additional $t_R \rightarrow b_L $ weak-moment coupling of
arbitrary relative strength $\Lambda_{+}$ is $ W_\mu ^{*} J_{\bar
b t}^\mu = W_\mu ^{*}\bar u_{b}\left( p\right) \Gamma ^\mu u_t
\left( k\right) $ where $k_t =q_W +p_b $, and
\begin{equation}
\frac{1}{2} \Gamma ^\mu =g_L\gamma ^\mu P_L + \frac{g_{f_M + f_E}
} {2\Lambda _{+} }\iota \sigma ^{\mu \nu } (k-p)_\nu P_R
\end{equation}
where $P_{L,R} = \frac{1}{2} ( 1 \mp \gamma_5 ) $.  In $g_L =
g_{f_M + f_E} = 1$ units, when $\Lambda_{+} = E_W/2$ which
corresponds to the (+) amplitudes, the complete $t \rightarrow b$
coupling is very simple
\begin{equation}
\gamma ^\mu P_L + \iota \sigma ^{\mu \nu } v_\nu P_R
 =P_R \left( \gamma
^\mu + \iota \sigma ^{\mu \nu } v_\nu \right)
\end{equation}
where $v_{\nu}$ is the W-boson's relativistic four-velocity.

In the  $t$ rest frame, the helicity-amplitude matrix element for
$t \rightarrow W^{+} b$ is \newline $ \langle \theta _1^t ,\phi
_1^t ,\lambda _{W^{+} } ,\lambda _b |\frac 12,\lambda _1\rangle =$
 $ D_{\lambda _1,\mu }^{(1/2)*}(\phi _1^t ,\theta _1^t ,0)A_{i}
\left( \lambda _{W^{+} } ,\lambda _b \right) $ where $\mu =\lambda
_{W^{+} } -\lambda _b $ in terms of the $W^+$ and $b$-quark
helicities.  The asterisk denotes complex conjugation, the final
$W^{+}$ momentum is in the $\theta _1^t ,\phi _1^t$ direction, and
$\lambda_1$ gives the $t$-quark's spin component quantized along
the $z$ axis. $\lambda_1$ is also the helicity of the $t$-quark if
one has boosted, along the ``$-z$" direction, back to the $t$ rest
frame from the $(t \bar{t})_{cm}$ frame.  It is this boost which
defines the $z$ axis in the $t$-quark rest frame for angular
analysis [4].  Explicit expressions for the helicity amplitudes
associated with each ``$i$" coupling are listed in Sec. 2. We
denote by $\Gamma$ the partial-width for the $t \rightarrow W^{+}
b $ decay channel and by $\Gamma_{L,T}$ the partial-width's for
the sub-channels in which the $W^{+}$ is respectively
longitudinally, transversely polarized; $\Gamma = \Gamma_L
+\Gamma_T $.  Similarly, ${\Gamma_L}|_{\lambda_b = - \frac{1}{2}}
$ denotes the partial-width for the W-longitudinal sub-channel
with b-quark helicity $\lambda_b = - \frac{1}{2}$.

The intensity-ratio equivalence-theorem states, ``As consequence
of Lorentz-invariance,  for the $t \rightarrow W^{+} b $ decay
channel each of the four ratios ${\Gamma_L}|_{\lambda_b = -
\frac{1}{2}} / {\Gamma}$, ${\Gamma_T}|_{\lambda_b = - \frac{1}{2}}
/ {\Gamma}$, ${\Gamma_L}|_{\lambda_b =  \frac{1}{2}} / {\Gamma}$,
${\Gamma_T}|_{\lambda_b =  \frac{1}{2}} / {\Gamma}$, is identical
for the pure $(V-A)$ coupling and for the $(V-A) + (f_M + f_E)$
coupling with $\Lambda_{+} =E_W /2$, and their respective
partial-widths are related by $\Gamma_{+} = v^2 \Gamma_{SM}$. $v$
is the velocity of the W-boson in the t-quark rest frame." Note
that this theorem does not require specific values of the mass
ratios $y \equiv m_W/m_t$, and $x \equiv m_b/m_t$, but that by
setting $\Lambda_{+} =E_W /2$ the relative strength of the chiral
weak-transition moment for the top quark has been fixed versus
$g_L$.

The three tWb-transformations, $A_{+}=M $ $A_{SM}$, $P $ $A_{SM}$,
$B $ $A_{SM}$, are related to this theorem.  The $M$
transformation implies the theorem, but as explained below, $M$
also implies the sign and ratio differences of the (ii) and (iii)
type amplitude ratio-relations which distinguish the (SM) and (+)
couplings. The $P$ and $B$ transformations more completely exhibit
the underlying symmetries relating these two Lorentz-invariant
couplings. In particular, these three $4$x$4$ matrices lead to two
``commutator plus anti-commutator" symmetry algebras, and together
enable a characterization of the values of $\Lambda_{+}$, $y
\equiv m_W/m_t$, and $x \equiv m_b/m_t$.  In Sec. 2, it is shown
how these three tWb-transformations successively arise from
consideration of different types of ``helicity amplitude
relations" for $t \rightarrow W^{+} b $ decay: The type (i) are
ratio-relations which hold separately for the two cases,
``$i=(SM)$, $(+)$". The type (ii) are ratio-relations which relate
the amplitudes in the two cases.  By the type (iii)
ratio-relations, the tWb-transformation $A_{+}=M$ $A_{SM}$ where
$M=v$ $diag(1,-1,-1,1)$ characterizes the mass scale $\Lambda_{+}
= E_W/2 $. Similarly, the amplitude condition (iv)
\begin{equation}
A_{+} (0,-1/2) = a A_{SM} (-1,-1/2),
 \end{equation}
with $a= 1 + O(v \neq y \sqrt{2}, x)$, determines the scale of the
tWb-transformation matrix $P$ and determines the value of the mass
ratio $y \equiv m_W/m_t$. $O(v \neq y \sqrt{2}, x)$ denotes small
corrections, see below. The amplitude condition (v)
\begin{equation}
A_{+} (0,-1/2) = - b A_{SM} (1-1/2),
\end{equation}
with $ b = v^{-8} $, determines the scale of $B$ and determines
the value of $ x= m_b/m_t$.  In Sec. 3, the two symmetry algebras
are obtained which involve the $M$, $P$, and $B$ transformation
matrices. Sec. 4 contains a discussion of these results and their
implications assuming that the observed $t \rightarrow W^{+} b $
decay mode will be found empirically to be well-described by (2).

\section{Helicity amplitude relations:}

In the Jacob-Wick phase convention, the helicity amplitudes for
the most general Lorentz coupling are given in [4]. In $g_L =
g_{f_M + f_E} = 1$ units and suppressing a common overall factor
of $\sqrt{m_t \left( E_b +q_W \right) }$, for only the $(V-A)$
coupling the associated helicity amplitudes are:
\begin{eqnarray*}
 A_{SM} \left(
0,-\frac 12\right) & = & \frac{1 }{y } \; \frac{E_W +q_W }{m_t }
\\
 A_{SM} \left(
-1,-\frac 12\right) & = & \sqrt{2}  \\
 A_{SM} \left( 0,\frac 12\right)
& = & -  \frac{1 }{y }  \frac{E_W -q_W }{m_t } \left( \frac
{m_b}{m_t-E_W +  q_W}  \right) \\
 A_{SM} \left( 1,\frac 12\right) & = & -
\sqrt{2} \left( \frac {m_b}{m_t-E_W +  q_W}  \right)
\end{eqnarray*}
For only the $(f_M + f_E)$ coupling, i.e. only the additional $t_R
\rightarrow b_L $ tensorial coupling:
\begin{eqnarray*}
 A_{f_M + f_E} \left(
0,-\frac 12\right) & = &  - ( \frac{m_t }{2\Lambda_+ }) \;  y  \\
 A_{f_M + f_E} \left(
-1,-\frac 12\right) & = & - ( \frac{m_t }{2\Lambda_+ }) \sqrt{2}
\; \frac{E_W +q_W }{m_t } \\
 A_{f_M + f_E} \left( 0,\frac 12\right)
& = & ( \frac{m_t }{2\Lambda_+ }) y \left( \frac {m_b}{m_t-E_W +
q_W}  \right) \\
 A_{f_M + f_E} \left( 1,\frac 12\right) & = & ( \frac{m_t }{2\Lambda_+ })
\sqrt{2} \;  \frac{E_W - q_W }{m_t } \left( \frac {m_b}{m_t-E_W +
 q_W}  \right)
\end{eqnarray*}
From these, the amplitudes for the $(V-A) + (f_M + f_E)$ coupling
of (1) are obtained by \newline $A_{+} (\lambda_W, \lambda_b) =
A_{SM} (\lambda_W, \lambda_b) + A_{f_M + f_E } (\lambda_W,
\lambda_b)$. For $\Lambda_{+} = E_W /2$, the $A_{+} (\lambda_W,
\lambda_b)$ amplitudes corresponding to the complete $t
\rightarrow b$ coupling (2) are
\begin{eqnarray*}
 A_{+} \left(
0,-\frac 12\right) & = & \frac{1 }{y } \; (q/E_W)  \; \frac{E_W
+q_W }{m_t }
\\
 A_{+} \left(
-1,-\frac 12\right) & = & - \sqrt{2} \; (q/E_W) \\
 A_{+} \left( 0,\frac 12\right)
& = & \frac{1 }{y } \; (q/E_W) \;  \frac{E_W -q_W }{m_t } \left(
\frac {m_b}{m_t-E_W +  q_W}  \right) \\
 A_{+} \left( 1,\frac 12\right) & = & -
\sqrt{2} \; (q/E_W) \; \left( \frac {m_b}{m_t-E_W +  q_W}  \right)
\end{eqnarray*}
For the three ``$i$" couplings, a direct derivation from (1) shows
how the different factors arise in the amplitudes [8].

We now analyze the different types of helicity amplitude relations
involving both the SM's amplitudes and those in the case of the
$(V-A) + (f_M + f_E)$ coupling: The first type of ratio-relations
holds separately for $i=(SM)$, $(+)$ and for all $y = \frac {m_W}
{m_t} , x = \frac {m_b} {m_t} , \Lambda_{+}$ values, (i):
\begin{equation}
\frac{A_{i} (0,1/2) } { A_{i} (-1,-1/2) } = \frac{1}{2}
\frac{A_{i} (1,1/2) } { A_{i} (0,-1/2) }
\end{equation}

The second type of ratio-relations relates the amplitudes in the
two cases and also holds for all $y, x, \Lambda_{+}$ values. The
first two relations have numerators with opposite signs and
denominators with opposite signs, c.f. Table 1; (ii): Two
sign-flip relations
\begin{equation}
\frac{A_{+} (0,1/2) } { A_{+} (-1,-1/2) } = \frac{A_{SM} (0,1/2) }
{ A_{SM} (-1,-1/2) }
\end{equation}
\begin{equation}
\frac{A_{+} (0,1/2) } { A_{+} (-1,-1/2) } = \frac{1}{2}
\frac{A_{SM} (1,1/2) } { A_{SM} (0,-1/2) }
\end{equation}
and two non-sign-flip relations
\begin{equation}
\frac{A_{+} (1,1/2) } { A_{+} (0,-1/2) } = \frac{A_{SM} (1,1/2) }
{ A_{SM} (0,-1/2) }
\end{equation}
\begin{equation}
\frac{A_{+} (1,1/2) } { A_{+} (0,-1/2) } = 2 \frac{A_{SM} (0,1/2)
} { A_{SM} (-1,-1/2) }
\end{equation}
(7, 9), which are not in [2], are essential for obtaining the $P$
and $B$ tWb-transformations and thereby the symmetry algebras of
Sec. 3 below.

The third type of ratio-relations, holding for all $y, x$ values,
follows by determining the effective mass scale, $\Lambda_{+}$, so
that there is an exact equality for the ratio of left-handed
amplitudes (iii):
\begin{equation}
\frac{A_{+} (0,-1/2) } { A_{+} (-1,-1/2) } = -
 \frac{A_{SM} (0,-1/2) } { A_{SM} (-1,-1/2) },
\end{equation}
Equivalently, $ \Lambda_{+} = \frac{m_t } {4 } [ 1 + (m_W / m_t)^2
- (m_b / m_t)^2] = E_W/2$ follows from each of:
\begin{equation}
\frac{A_{+} (0,-1/2) } { A_{+} (-1,-1/2) } = -
 \frac{1}{2} \frac{A_{SM} (1,1/2) } { A_{SM} (0,1/2) },
\end{equation}
\begin{equation}
\frac{A_{+} (0,1/2) } { A_{+} (1,1/2) } = - \frac{A_{SM} (0,1/2) }
{ A_{SM} (1,1/2) },
\end{equation}
\begin{equation}
\frac{A_{+} (0,1/2) } { A_{+} (1,1/2) } = - \frac{1}{2}
\frac{A_{SM} (-1,-1/2) } { A_{SM} (0,-1/2) },
\end{equation}
From the amplitude expressions given above, the value of this
scale $\Lambda_{+}$ can be characterized by postulating the
existence of a tWb-transformation $A_{+}=M$ $A_{SM}$ where $M=v$
$diag(1,-1,-1,1)$, with $
A_{SM}=[A_{SM}(0,-1/2),A_{SM}(-1,-1/2),A_{SM}(0,1/2),A_{SM}(1,1/2)]$
and analogously for $A_{+}$.

Assuming (iii), the fourth type of relation is the equality (iv):
\begin{equation}
A_{+} (0,-1/2) = a A_{SM} (-1,-1/2),
\end{equation}
where $a= 1 + O(v \neq y \sqrt{2}, x)$.

This is equivalent to the velocity formula $  v = a  y \sqrt{2}
\left( \frac{1} {1- ( E_b - q_W )/m_t} \right)
 = a y \sqrt{2} $ for $ m_b = 0 $.
In [2], for $a=1$ it was shown that (iv) leads to a mass relation
with the solution $y=\frac {m_W} {m_t} = 0.46006$ ($ x=0$). The
present empirical value is $ y = 0.461 \pm 0.014$, where the error
is dominated by the $3 \% $ precision of $m_t$. In [2], for $a=1$
it was also shown that (iv) leads to $\sqrt{2}=v\gamma
(1+v)=v\sqrt{\frac{1+v}{1-v}}$ so $v=0.6506\ldots$ without input
of a specific value for $m_b$. However, by Lorentz invariance $v$
must depend on $m_b$. Accepting (iii), we interpret this to mean
that $a \neq 1$ and in the Appendix obtain the form of the $O(v
\neq y \sqrt{2}, x)$ corrections in $a$ as required by Lorentz
invariance.  The small correction $O(v \neq y \sqrt{2}, x)$
depends on both $x \equiv m_b/m_t$ and the difference $v-y
\sqrt{2}$.

Equivalently, by use of (i)-(iii) relations, (14) can be expressed
postulating the existence of a second tWb-transformation $A_{+}=P
$ $A_{SM}$ where
\begin{equation}
P\equiv v
\left[
\begin{array}{cccc}
0 & a/v & 0 & 0 \\ -v/a & 0 & 0 & 0 \\ 0 & 0 & 0 & -v/2a \\ 0 & 0
& 2a/v & 0
\end{array}
\right]
\end{equation}
The value of the parameter $a$ of (iv) is not fixed by (15).

The above two tWb-transformations do not relate the $\lambda_b= -
\frac{1}{2}$ amplitudes with the $\lambda_b= \frac{1}{2}$
amplitudes. From (i) thru (iv), in terms of a parameter $b$, the
equality (v):
\begin{equation}
A_{+} (0,-1/2) = - b A_{SM} (1,1/2),
\end{equation}
is equivalent to $A_{+}=B $ $A_{SM}$
\begin{equation}
B\equiv v \left[
\begin{array}{cccc}
0 & 0 & 0 & -b/v \\ 0 & 0 & 2b/v & 0 \\ 0 & v/2b & 0 & 0
\\ -v/b & 0 & 0 & 0
\end{array}
\right]
\end{equation}
The choice of $ b = v^{-8} = 31.152$, gives
\begin{equation} B\equiv v
\left[
\begin{array}{cccc}
0 & 0 & 0 & -v^{-9} \\ 0 & 0 & 2v^{-9} & 0 \\ 0 & v^{9}/2 & 0 & 0
\\ -v^{9} & 0 & 0 & 0
\end{array}
\right]
\end{equation}
and corresponds to the mass relation $ m_{b}
=\frac{m_{t}}{b}\left[ 1-\frac{vy}{\sqrt{2}}\right] =4.407...GeV $
for $m_t = 174.3 GeV$.

\section{Commutator plus anti-commutator symmetry algebras:}

The anti-commuting matrices
\begin{equation}
m\equiv \left[
\begin{array}{cc}
1 & 0 \\ 0 & -1
\end{array}
\right] ,p\equiv \left[
\begin{array}{cc}
0 & -a/v \\ v/a & 0
\end{array}
\right] ,q\equiv \left[
\begin{array}{cc}
0 & a/v \\ v/a & 0
\end{array}
\right]
\end{equation}
satisfy $[m,p]=-2q,[m,q]=-2p,[p,q]=-2m$. \ Similarly, $m$ and
\begin{equation}
r\equiv \left[
\begin{array}{cc}
0 & -v/2a \\ 2a/v & 0
\end{array}
\right] ,s\equiv \left[
\begin{array}{cc}
0 & v/2a \\ 2a/v & 0
\end{array}
\right]
\end{equation}
are anti-commuting and satisfy $%
[m,r]=-2s,[m,s]=-2r,[r,s]=-2m$. Note $m^2=q^2=s^2=1$,
$p^2=r^2=-1$, and that $a$ is arbitrary.  Consequently, if one
does not distinguish the $(+)$ versus SM indices, respectively of
the rows and columns, the tWb-transformation matrices have some
simple properties:

The anticommuting 4x4 matrices
\begin{equation}
M\equiv v\left[
\begin{array}{cc}
m & 0 \\ 0 & -m
\end{array}
\right] ,P\equiv v\left[
\begin{array}{cc}
-p & 0 \\ 0 & r
\end{array}
\right] ,Q\equiv v\left[
\begin{array}{cc}
q & 0 \\ 0 & s
\end{array}
\right]
\end{equation}
satisfy the closed algebra
$[\overline{M},\overline{P}]=2\overline{Q},[\overline{M},\overline{Q}]
=2\overline{P},[\overline{P},\overline{Q}]=2\overline{M}$. The bar
denotes removal of the overall ``$v$" factor, $M= v \overline{M},
...$.  Note that $Q$ is not a tWb-transformation.

Including the B matrix with $b$ arbitrary, the algebra closes with
3 additional matrices
\begin{equation}
\overline{B}\equiv \left[
\begin{array}{cc}
0 & d \\ f & 0
\end{array}
\right] ,\overline{C}\equiv \left[
\begin{array}{cc}
0 & e \\ g & 0
\end{array}
\right]
\end{equation}

\begin{equation}
\overline{G}\equiv \left[
\begin{array}{cc}
0 & h \\ k & 0
\end{array}
\right] ,\overline{H}\equiv \left[
\begin{array}{cc}
0 & j \\ l & 0
\end{array}
\right]
\end{equation}
where
\begin{equation}
d\equiv \left[
\begin{array}{cc}
0 & -b/v \\ 2b/v & 0
\end{array}
\right] ,e\equiv \left[
\begin{array}{cc}
0 & b/v \\ 2b/v & 0
\end{array}
\right] ,f\equiv \left[
\begin{array}{cc}
0 & v/2b \\ -v/b & 0
\end{array}
\right] ,g\equiv \left[
\begin{array}{cc}
0 & v/2b \\ v/b & 0
\end{array}
\right]
\end{equation}

\begin{equation}
h\equiv \left[
\begin{array}{cc}
-2ab/v^{2} & 0 \\ 0 & b/a
\end{array}
\right] ,j\equiv \left[
\begin{array}{cc}
2ab/v^{2} & 0 \\ 0 & b/a
\end{array}
\right] ,
 \newline
k\equiv \left[
\begin{array}{cc}
1/2v^{2}ab & 0 \\ 0 & -a/b
\end{array}
\right] ,l\equiv \left[
\begin{array}{cc}
1/2v^{2}ab & 0 \\ 0 & a/b
\end{array}
\right]
\end{equation}
The squares of the $2$x$2$ matrices (24-25) do depend on $a$, $b$,
and $v$.

The associated closed algebra is: $[\overline{M},\overline{B}]=0,\{\overline{%
M},\overline{B}\}=-2\overline{C};[\overline{B},\overline{C}]=0,\{\overline{B}%
,\overline{C}\}=-2\overline{M};$ \newline
$[\overline{M},\overline{C}]=0,\{\overline{M},%
\overline{C}\}=-2\overline{B};$ and
$[\overline{P},\overline{B}]=2\overline{H}
,\{\overline{P},\overline{B}\}=0;[%
\overline{H},\overline{P}]=2\overline{B},\{\overline{H},\overline{P}\}=0;$
\newline
$[%
\overline{H},\overline{B}]=2\overline{P},\{\overline{H},\overline{B}\}=0$
. Similarly,  $[\overline{P},\overline{C}]=0,\{\overline{P},\overline{%
C}\}=-2\overline{G};[\overline{M},\overline{H}]=-2\overline{G},$
\newline
$ \{\overline{M}%
,\overline{H}\}=0;[\overline{H},\overline{C}]=0,\{\overline{H},\overline{C}%
\}=2\overline{Q};$ and
$[\overline{M},\overline{G}]=-2\overline{H},\{%
\overline{M},\overline{G}\}=0;[\overline{P},\overline{G}]=0,$
\newline
$\{\overline{P},\overline{G}\}=2\overline{C};[\overline{G},\overline{B}]=-2%
\overline{Q},\{\overline{G},\overline{B}\}=0;$ and  $[\overline{G},\overline{C%
}]=0,\{\overline{G},\overline{C}\}=-2\overline{P};$ \newline $[\overline{G},\overline{H}%
]=2\overline{M},\{\overline{G},\overline{H}\}=0.$ The part involving  $%
\overline{Q}$ \  is
$[\overline{G},\overline{Q}]=2\overline{B},\{\overline{G},\overline{Q}\}=0;[%
\overline{B},\overline{Q}]=2\overline{G},$ \newline
 $ \{\overline{B},\overline{Q}\}=0;
 [\overline{C},\overline{Q}]=0,
 \{\overline{C},\overline{Q}\}=-2\overline{H};$ $%
[\overline{H},\overline{Q}]=0,\{\overline{H},\overline{Q}\}=
2\overline{C}$.

\bigskip

\ This has generated an additional tWb-transformation $G\equiv v\overline{G}$%
; but $C\equiv v\overline{C}$ and $H\equiv v\overline{H}$ are not
tWb-transformations. \

\bigskip

Up to the insertion of an overall $\iota =\sqrt{-1}$, each of
these 4x4 barred matrices is a resolution of unity, i.e.
$\overline{P}^{-1}=-\overline{P}$,
$\overline{G}^{-1}=-\overline{G}$, but
$\overline{Q}^{-1}=\overline{Q}$,
$\overline{B}^{-1}=\overline{B},...$ .

\section{Discussion:}

\indent {\bf (1)  $\Lambda_{+}$ mass scale:}

A fundamental question [2] raised by the existence of the
intensity-ratio equivalence theorem is ``What is the origin of the
$\Lambda_{+} =E_W /2 \sim 53 GeV$ mass scale?"  The present paper
shows that $M$ is but one of three logically-successive
tWb-transformations which are constrained by the helicity
amplitude ratio-relations (i) and (ii). Thereby, the type (iii)
ratio-relation fixes $\Lambda_{+} = E_W/2$ and the overall scale
of the tWb-transformation matrix $M$. The amplitude condition
(iv), $A_{+} (0,-1/2) = a A_{SM} (-1,-1/2)$ with $a= 1 + O(v \neq
y \sqrt{2}, x)$, and the amplitude condition (v), $A_{+} (0,-1/2)
= - b A_{SM} (1-1/2)$ with $ b = v^{-8} $, determine respectively
the scale of the tWb-transformation matrices $P$ and $B$ and
characterize the values of $m_W/m_t$ and $m_b/m_t$. The overall
scale can be set here by $ m_t $ or $ {m_W}$. From an empirical
``bottom-up" perspective of further ``unification", $ m_W$ is more
appropriate to use to set these scales since its value is fixed in
the SM by the vacuum expectation value of the Higgs field, $\phi$
[ the SM's EW scale is $v_{EW}=\sqrt{-\mu^2 / \vert \lambda \vert}
= \sqrt{2} \langle 0|\phi |0\rangle \sim 246GeV$].  So by
postulating the $M$ transformation, the $\Lambda_{+}$ mass scale
is fixed as $E_W/2$. By these three tWb-transformations, the
numerical value of $\Lambda_{+} $ is determined by that of
$v_{EW}$.

{\bf (2)  Comparison with an Amplitude Equivalence-Theorem:}

Given the continued successes of predictions based on the
couplings and symmetries built into the SM, and given the present
rather slow pace of new experimental information, we appreciate
the fact that for many readers it can be difficult to remind
oneself that directly from experiment we still really do not know
much about the properties of the on-shell top-quark [1]. Because
of possible form factor effects and possible unknown thresholds
due to new particles, in a theory/model-independent manner one
cannot reliably determine the indirect constraints on on-shell
top-quark couplings from off-shell contributions from top-quark
contributions in higher-order loops in electroweak precision tests
[2].  Various assumptions of quark universality are still
routinely made in the theoretical literature to conjecture
on-shell top-quark properties from theoretical patterns found for
the several-order-of-magnitude less massive quarks, in spite of
the closeness of the value of $m_t$ to $v_{EW}$ and of the now
significantly greater numerical-precision ($3\%$) of the $m_t$
measurement [1] than that of any of the other quark masses.

This present ``not-knowing" status quo is very different from that
in 1940-1952 in regard to the then rapidly changing and developing
experimental status quo in the case of the weak and strong
interactions, which was concurrent with the series of striking
empirical-theoretical successes with QED ( QED is often viewed as
the prototypical earlier analogue of the present SM ).
Nevertheless, despite these differences in the experimental
situation, we think it is instructive to compare this present
intensity-ratio equivalence theorem with a somewhat analogous
``amplitude equivalence theorem (ET)" which was discovered and
quite intensely studied, circa 1940-1952, for the pseudoscalar and
pseudovector interactions, Dyson (1948) [9].

Although not as influential as Fermi's 1934 paper and the
Gamow-Teller 1936 paper concerning the Lorentz structure of the
weak interactions, this ET has had a long, fruitful, and
significant impact in high energy experiment and in theoretical
elementary particle physics.  The early history and early
significance of the ET , e.g.  [9] and [10], can be traced from
Schweber's book (1962) [11] and from the entire final chapter of
Schweber, Bethe, and de Hoffmann, (1955) [12]. This ET stimulated
in part, the development of effective Lagrangian methods [13] and
work by G. t'Hooft and M. Veltman [14].  ( Later ET literature
cites the t'Hooft-Veltman preprint; the corresponding published
tHV-paper does not cite, e.g. [12], but refers to their tHV-paper
as being based on unpublished preprints. ) The ``theorem"
continues to be of theoretical interest, see for instance [15]. In
top-quark physics, we think it is reasonable to expect that the
theoretical patterns, analytic relationships and
tWb-transformations developed in the present paper between the
helicity amplitudes  for this specific additional tensorial
coupling and those for the $(V-A)$ coupling will have such an
early experimental-theoretical history.

There are other similarities and other important qualitative
differences between the present situation and that concerning the
ET.  They are similar in that both relate simple Lorentz
structures, involve helicities...though more intricately in the
present case, and involve a renormalizable coupling. Major
differences include (I) the ET case is much better understood
after over 60 years of research papers, whereas this is only a 2nd
paper towards understanding the patterns and possible physics of
this aspect of top quark decays, (II) the SM is now known to well
explain most of the weak interaction systems (nucleon, nucleus,
strange particle weak decays) first studied by the ET stimulating
experiments, whereas experiments have only begun on top quark
decay, and (III) the tWb transformations involve couplings of a
fundamental renormalizable local quantum field theory, the SM, and
fundamental mass ratios, whereas we now know that the ET case
never did.

{\bf (3) Possible Implications of These Symmetries:}

The additional $t_R \rightarrow b_L$ weak-moment coupling violates
the conventional gauge invariance transformations of the SM and
traditionally in electroweak studies such anomalous couplings have
been best considered as ``induced" or ``effective". Nevertheless,
in special ``new physics" circumstances such a simple
charge-changing tensorial coupling as (2) might turn out to be a
promising route to deeper understandings. The ``tensorial
coupling" is a basic structure if considered from gravitation
viewpoints. However, are the new symmetries associated with the
symmetry algebras of Sec. 3 sufficient to overcome the known
difficulties [16] in constructing a renormalizable, unitary
quantum field theory involving second class currents [17] ? The
$f_E$ component is second class. $f_E$ has a distinctively
different reality structure, and time-reversal invariance property
versus the first class $ V, A, f_M$ [18].  If the observed $t
\rightarrow W^{+} b $ decay mode is found empirically to be well
described by (2) this would support a working hypothesis that in
the renormalization of the $ t \rightarrow b$ coupling and of the
top and bottom quark masses, the underlying symmetries of these
tWb-transformations are basic to relating the associated mass
scales, much as are Lorentz invariance and the symmetries of
gauge-invariance-dynamics in performing renormalizations in the
SM.

{\bf (4)  Experimental Tests/Measurements:}

In on-going [1] and forth-coming [6,7] top-quark decay
experiments, important information about the relationship of the
tWb-transformation symmetry patterns of this paper to the observed
top quark decays will come from:
\newline (a) Measurement of the sign of the $\eta_L \equiv \frac
1\Gamma |A(-1,-\frac 12)||A(0,- \frac 12)|\cos \beta _L = \pm
0.46(SM/+) $ helicity parameter [4] so as to determine the sign of
$cos \beta_L $ where $ \beta_L = \phi _{-1}^L- \phi _0^L $ is the
relative phase of the two $\lambda_b = - \frac{1}{2} $ amplitudes,
$A(\lambda _{W^{+}},\lambda _b)=|A|\exp (i\phi _{\lambda
_{W^{+}}}^{L,R})$.  For the exclusion of the coupling of (2)
versus the SM's $(V-A)$ coupling, this would be the definitive
near-term measurement concerning properties of the on-shell
top-quark.
\newline
 (b) Measurement, or an empirical bound, for the
closely associated \newline $ {\eta_L}^{'} \equiv \frac 1\Gamma
|A(-1,-\frac 12)||A(0,- \frac 12)|\sin \beta _L $ helicity
parameter.  This would provide useful complementary information,
since in the absence of $T_{FS}$-violation, ${\eta_L}^{'} =0$ [4].
\newline
 (c) Measurement of the partial width for $t \rightarrow
W^+ b$ such as in single-top production [19-21]. The $v^2$ factor
which differs their associated partial widths corresponds to the
SM's $\Gamma_{SM}= 1.55 GeV$, versus $\Gamma_+ = 0.66 GeV$ and a
longer-lived (+) top-quark if this mode is dominant.

    Since the helicity amplitude relations discussed in Sec. 2 involve
the b-quark helicities, c.f. differing signs in $\lambda_b = 1/2$
column of Table 1, there are also independent phase tests which
require
\newline
  (d) Measurements of helicity parameters [5] using $\Lambda_b$-polarimetry in stage-two
spin-correlation functions.  It is noteworthy that the $\Lambda_b$
baryon has been observed by CDF at the Tevatron [22].

{\bf Acknowledgments: }

We thank experimental and theoretical physicists for discussions.
This work was partially supported by U.S. Dept. of Energy Contract
No. DE-FG 02-86ER40291.

{\bf Appendix: The $O(v \neq y \sqrt{2}, x)$ corrections in $a$}

In this appendix is listed the form of the $O(v \neq y \sqrt{2},
x)$ corrections in $a$ as required by Lorentz invariance:

For $a=1+\varepsilon (x,y)$, the (iv) relation is $v=(1+\varepsilon )y \sqrt{2}%
m_{t}/(E_{W}+q)$ whereas from relativistic kinematics $\ \
v=q/E_{W}=[(1-y^{2}-x^{2})^{2}-4y^{2}x^{2}]^{1/2}/[1+y^{2}-x^{2}]$
. \ By equating these expressions and expanding in $x$, one
obtains $ \varepsilon =R+x^{2}S $ where
\begin{eqnarray*}
R &=&\frac{1-4y^{2}-3y^{4}-2y^{6}}{4y^{2}(1+y^{2})^{2}} \\ S
&=&\frac{-1-4y^{2}+y^{4}}{2y^{2}(1+y^{2})^{3}}
\end{eqnarray*}
and $  v= y \sqrt{2} \left[
1+R+x^{2}(S+\frac{1+R}{1-y^{2}})+O(x^{4})\right] $.  From the
latter equation, $R=(v-y\sqrt{2})/y\sqrt{2}+O(x^{2}).$

For a massless b-quark ( $x=0$ ) and $a=1$ , the (iv) relation is
equivalent
to the $\frac{m_{W}}{m_{t}}$ mass relation \  $y^{3}\sqrt{2}$ $+y^{2}+y\sqrt{%
2}-1=0$, and by relativistic kinematics to the W-boson velocity condition $%
v^{3}+v^{2}+2v-2=0$ and the simple formula $v=y\sqrt{2}$.

\newpage

\begin{center}
{\bf Table Captions}
\end{center}

Table 1:  Numerical values of the helicity amplitudes for the
standard model $(V-A)$ coupling and for the (+) coupling of
Eq.(2). The latter consists of an additional $t_R \rightarrow b_L$
weak-moment of relative strength $\Lambda_{+} \sim 53 GeV$ so as
yield a relative-sign change in the $\lambda_b= - \frac{1}{2}$
amplitudes. The values are listed first in $ g_L = g_{f_M + f_E} =
1 $ units, and second as $ A_{new} = A_{g_L = 1} / \surd \Gamma $.
Table entries are for $m_t=175GeV, \; m_W = 80.35GeV, \; m_b =
4.5GeV$.


\begin{thebibliography}{333}

\bibitem{fnal} CDF collaboration, Affolder T  2000 {\it et al Phys Rev Let } {\bf 84},
216; D{\O} collaboration, Abbott B 2000 {\it et al Phys Rev Let }
{\bf 85} 256; and Chakraborty D, Konigsberg J and Rainwater D 2003
{\it Ann Rev Nucl Part Sci} {\bf 53}  301.
\bibitem{one}  Nelson C A 2002 {\it Phys Rev } {\bf D65} 074033;
 2001 p. 369 {\it Physics at Extreme Energies}
eds. Nguyen van Hieu and Jean Tran Thanh Van (Gioi Publishers,
Vietnam).
\bibitem{JW} Jacob M and Wick G 1959 {\it Ann Phys }{\bf 7} 209.
\bibitem{nklm1} Nelson C A, Kress B T, Lopes M and McCauley T P 1997 {\it Phys
Rev} {\bf D56} 5928; 1998 {\it ibid} {\bf D57} 5923; Nelson C A
and Cohen A M 1999 {\it Eur Phys J} {\bf C8} 393; Nelson C A and
Adler L J 2000 {\it ibid} {\bf C17} 399.
\bibitem{lambdab} Nelson C A 2001 {\it Eur Phys J} {\bf C19}
323.
\bibitem{LHC} ATLAS Technical Proposal 1994, CERN/LHCC/94-43,
LHCC/P2; CMS Technical Design Report 1997, CERN-LHCC- 97-32;
CMS-TDR-3.
\bibitem{nlc} APS/DPF/DPB Summer Study, Snowmass-2001-E3020,
eConf C010630; Tesla Tech. Design Report, DESY 2001-011,
http://tesla.desy.de/; Japanese Linear Collider Group, JLC-I,
KEK-Report 92-16(1992); Linear Collider Physics Resource Book,
American Linear Collider Working Group, SLAC-R-570,
http://www.slac.stanford.edu/grp/th/LCBook/.
\bibitem{Kim}  Kim N H 2003 {\it Thesis, SUNY at Binghamton}.
\bibitem{cdyson} Dyson F J 1948 {\it Phys Rev} {\bf 73} 929;
Nelson E C 1941 {\it ibid} {\bf 60} 830.
\bibitem{et} Case K M 1949 {\it Phys Rev} {\bf 76} 1; {\it ibid}
14; Wentzel G 1952 {\it ibid} {\bf 86} 802; Berger J M, Foldy L L
and Osborn R K 1952 {\it ibid} {\bf 87} 1061; Drell S D and Henley
E M 1952 {\it ibid} {\bf 88} 1053.
\bibitem{schs} Schweber S S {\it An Introduction to Relativistic
 Quantum Field Theory} , p. 301-7, (Harper \& Row, New York 1962).
\bibitem{sbdh} Schweber S S, Bethe H A and de Hoffmann F {\it Mesons and
Fields}, Vol. 2, Chap. 26 ( Row, Peterson and Co, Evanston, IL
1955).
\bibitem{41} Coleman S, Wess J and Zumino B 1969 {\it Phys
Rev} {\bf 177} 2239.
\bibitem{42} t'Hooft G and Veltman M 1972 {\it Combinatorics of
Gauge Fields, preprint} (Utrecht, 1972); 1972 {\it Nuc Phys } {\bf
B50} 318.
\bibitem{43} Gomis J and Weinberg S 1996 {\it Are
nonrenormalizable gauge theories renormalizable?, Nuc Phys}  {\bf
B469} 473(1996); Ferrari R, Picariello M and Quadri A 2002 {\it
JHEP} {\bf 204} 33.
\bibitem{c7} Langacker P 1976 {\it Phys Rev} {\bf D14} 2340; 1977 {\bf
D15} 2386.
\bibitem{c8} Weinberg S 1958 {\it Phys Rev} { \bf 112} 1375.
\bibitem{c9}  Cabbibo N 1964 {\it Phys Rev Lett} {\bf 12} 137;
Nelson C A 1995 {\it Phys Let} {\bf B355} 561.
\bibitem{st1} Willenbrock S and Dicus D A 1986 {\it Phys Rev} {\bf D34}
155.
\bibitem{st2} Boos E, Dudko L and Ohl T 1999 {\it Eur Phys J}
{\bf C11} 473); Boos E, Dubinin M, Sachwitz M, and Schreiber H J
2001 {\it Eur Phys J } {\bf C21} 81; and Kauer N and Zeppenfeld D
2002 {\it ibid} { \bf D65} 014021.
\bibitem{n1} del Aguila F and Aguilar-Saavedra J A 2003 {\it Phys Rev
} {\bf D67} 014009; Espriu D and Manzano J 2002 {\it ibid} {\bf
D66} 114009.
\bibitem{Bphysics} Paulini M hep-ex/0402020.

\end{thebibliography}
\end{document}